\documentclass{llncs}
\usepackage{llncsdoc}

\usepackage{amsmath}
\usepackage{amssymb}
\usepackage{graphicx}
\usepackage{color}
\usepackage{array}
\newcolumntype{C}[1]{>{\centering\let\newline\\\arraybackslash\hspace{0pt}}m{#1}}

\usepackage[usenames,dvipsnames,table,svgnames]{xcolor}
\usepackage{xspace}
\usepackage{algpseudocode}

\usepackage{tikz}
\usetikzlibrary{arrows,positioning} 

\usepackage{enumerate,amsfonts,subfig}
\usepackage{verbatim,bm}
\usepackage{stmaryrd}
\usepackage{paralist}
\usepackage{pifont}
\usepackage{arydshln}

\usepackage{graphicx}   

\usepackage[breakable, theorems, skins]{tcolorbox}
\usepackage{listings}
\usepackage{soul}

\usepackage[T1]{fontenc}
\usepackage{float}
\usepackage{pgfplotstable}
\usepackage{pgfplots}
\usepackage{color}
\definecolor{bluekeywords}{rgb}{0.02,0.02,0.8}
\definecolor{greencomments}{rgb}{0.05,0.4,0.2}
\definecolor{redstrings}{rgb}{0.9,0,0}

\usepackage{multirow}
\usepackage{wrapfig}
\usepackage{listings}

\usepackage{enumitem}

\newcommand{\comm}[1]{}

\def\Name{\textsc{JDial}\xspace}
\def\Qlose{\textsc{Qlose}\xspace}
\def\Namep{\textsc{JDial+}\xspace}
\def\Qlosep{\textsc{Qlose+}\xspace}

\def\Namesingle{\textsc{JDial$_1$}\xspace}
\def\Namesingleopt{\textsc{JDial$_1^\text{o}$}\xspace}

\algtext*{EndWhile}
\algtext*{EndIf}
\algtext*{EndFor}
\algtext*{EndFunction}
\algtext*{skip}

\newcommand{\Prog}{P}

\newcommand{\inp}{i}

\newcommand{\inpbar}{I}

\newcommand{\varbar}{V}
\newcommand{\Loc}{L}

\newcommand{\val}{\nu}
\newcommand{\ns}{\textbf{?}}

\newcommand{\config}{\eta}
\newcommand{\exec}{\pi}

\newcommand{\repair}{\mathcal{RM}}
\newcommand{\progspace}{\mathcal{P}}

\newcommand{\naturalnumber}{\mathbb{N}}

\newcommand{\sketch}{\textsc{Sketch}\xspace}

\newcommand{\exit}{exit}

\newcommand{\test}{t}

\newcommand{\T}{T}

\renewcommand{\t}[1]{\texttt{#1}}

\newcommand{\xmark}{\ding{55}\xspace}%

\newcommand{\ind}{k}

\newcommand{\mypar}[1]{\vspace{1mm}\noindent\textit{#1}}
\newcommand{\bpar}[1]{\vspace{1mm}\noindent\textbf{#1.}}

\newcommand{\Quant}{Cost-aware}
\newcommand{\quant}{cost-aware}
\newcommand{\repspace}{repair space}
\newcommand{\trdist}{dist}
\newcommand{\tr}{\pi}

\newcommand{\rone}{(\emph{i})~}
\newcommand{\rtwo}{(\emph{ii})~}
\newcommand{\rthree}{(\emph{iii})~}

\definecolor{diffstart}{named}{Blue}
\definecolor{mark}{named}{Gray}
\definecolor{diffincl}{rgb}{0.05,0.4,0.2}
\definecolor{diffrem}{rgb}{0.6,0.1,0.1}

\lstdefinelanguage{diff}{
    basicstyle=\footnotesize\ttfamily,
    morecomment=[f][\color{diffstart}]{@@},
    morecomment=[f][\bfseries\color{mark}]{>},
    morecomment=[f][\color{diffincl}]{+\ },
    morecomment=[f][\color{diffrem}]{-\ },
    morecomment=[f][\color{greencomments}]{//\ },    
    morekeywords={def, while, return, if, for, return, int, void}
}

\begin{document}

\title{Program Repair via Direct State Manipulation}         


\author{Qinheping Hu$^1$, Isaac Evavold$^1$, Roopsha Samanta$^2$,\\Rishabh Singh$^3$, Loris D'Antoni$^1$}
\institute{$^1$ University of Wisconsin-Madison\\ $^2$ Purdue University \\ $^3$Microsoft}
%

\maketitle
\vspace{-10mm}
\begin{abstract}
The goal of program repair is to automatically fix programs to meet a 
specification.
We propose a new specification mechanism, 
\emph{direct manipulation}, in which
the  programmer can visualize the trace of a buggy program on a failing input 
and convey the intended program behaviour by manipulating variable values at some location.
The repair problem is to 
find a program that, on the same input, reaches 
the location identified by the programmer with variable values equal to the manipulated ones.  
Since a single program execution under-specifies the overall program behavior, we augment our repair problem with
quantitative objectives to find the program that agrees with the specification and is closest to the original one with respect to
some distance. 
We formalize the  repair problem, build 
a program repair tool \Name based on the  \sketch synthesizer, and 
show the effectiveness of \Name on representative 
buggy benchmarks from introductory programming assignments.
\end{abstract}

\section{Introduction}

The goal of program repair~\cite{repairsurvey} is to automatically fix buggy programs to meet a specification 
that is typically provided in the form of assertions or a set of test cases.
In this paper, we propose a new complementary specification mechanism called \emph{direct manipulation},
in which users can specify the intended program behavior by directly manipulating intermediate variable 
values in  buggy program traces.
We propose a workflow in which the user traverses the step-by-step visualization of 
the execution of the buggy program on a certain input to identify a location where the 
values of the program variables do not correspond to the ones she expects. 
At this point, we allow the user to \emph{manipulate} the variable values at the identified location
and modify them. 
We then treat this manipulation as a specification and use it to find a program that, 
on the same input, can reach the location identified by the user with the new variable values she provided.

We formalize our repair problem and present a 
constraint-based synthesis technique for computing program repairs based on direct manipulations. 
To repair programs in this setting we need to address two key challenges. 
First, the execution step manipulated by the user in the buggy trace might appear at a completely different point in the trace of the repaired program---e.g., when the repaired program uses a different number of loop iterations from the original program. 
Second, since a single program execution under-specifies the overall program behavior, there can be
many possible repaired programs that agree with the manipulated trace.
To address the first challenge,
given a manipulated location $\ell$,
we design an encoding that
 ``guesses'' in what occurrence of the location $\ell$ in the trace of the repaired program the desired variable values are produced. 
To address the second challenge, we augment our repair problem with quantitative objectives~\cite{qlose}
to prefer repairs that produce execution traces similar to those of the original program.

We implemented our synthesis technique
in a tool called \Name, which is built on top of the \textsc{Sketch}~\cite{sketch} synthesizer.
\Name supports several repair models and program distances, and
can repair Java programs containing loops, arrays, and recursion.
We also introduce an algorithm that uses concrete executions
to ``discover'' interpretations of external functions
and
allows \Name to repair programs containing library functions such as \texttt{Math.pow}.
Additinally, \Name also allows users to specify input-output test cases in addition to direct manipulations and
is therefore compatible with repair tools like \textsc{Qlose}~\cite{qlose}.

We evaluated \Name on representative benchmarks obtained from \textsc{Qlose}~\cite{qlose}
and several new benchmarks.
Our evaluation shows that \Name can quickly repair small Java programs and it produces more accurate repairs than 
techniques based only on test cases.
In particular, even when provided with a single input trace, \Name 
is able to compute correct repairs for 66\% of the programs while techniques
based on only test cases \emph{always} fail.

\bpar{Contributions} We make the following key contributions.
\begin{itemize}[leftmargin=*]
\item 
We introduce a specification mechanism called \emph{direct manipulation} 
together with a corresponding repair problem~(\S~\ref{sec:repair}).
\item 
We propose a modular framework based on sketching for repairing programs using direct manipulations (\S~\ref{sec:architecture}).

\item 
We instantiate our framework in \Name, a tool that can repair simple Java programs.
\Name supports a new algorithm
for repairing programs with external functions and an
optimized repair technique based on slicing
(\S~\ref{sec:implementation}).

\item 
We evaluate \Name on 17 representative benchmarks and show
\Name computes correct repairs in cases where  techniques
based on test cases fail (\S~\ref{sec:evaluation}).
\end{itemize}

\section{Illustrative Example}
\label{sec:motivating}

In this section, we illustrate our approach using an
example student attempt to an introductory programming exercise.
 In this domain, automatic program repair can be used to
provide personalized feedback 
to students~\cite{SGS13,Yi17,vlhcc17}.

\newsavebox\mybox
\begin{lrbox}{\mybox}
	 \scriptsize	\begin{tabular}{|lccc|}
	 		\hline
	 		& \t{max} & \t{min} & \t{i}  \\
	 		\hline
	 		\textit{value} &  5 & 4 & 1  \\
	 		\textit{change to}  	  & \textbf{9} & \ns & \ns \\
	 		\hline
	 	\end{tabular}
\end{lrbox}

\DeclareRobustCommand{\hlgray}[1]{{\sethlcolor{lightgray}\hl{#1}}}
\begin{figure*}[!t]
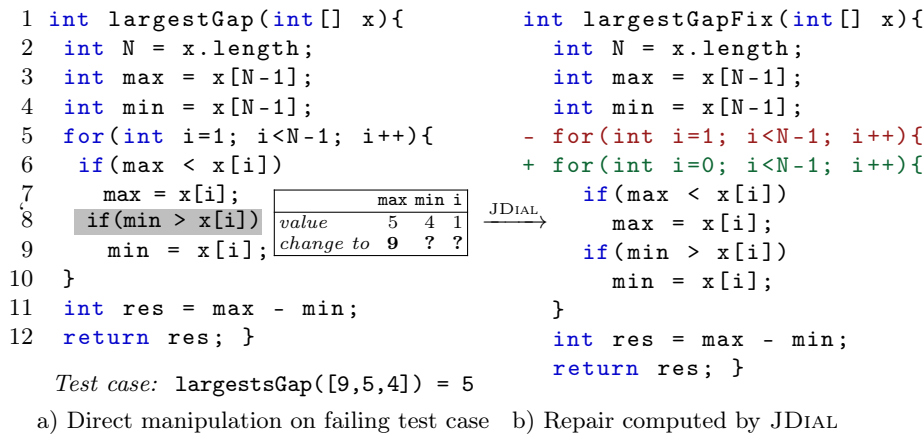
`
\centering
\begin{minipage}{.48\textwidth}
\centering
\begin{lstlisting}[
language=diff, numbers=left, numbersep=5pt, xleftmargin=.03\textwidth,
escapeinside={(*@}{@*)}
]
int largestGap(int[] x){
 int N = x.length;
 int max = x[N-1];
 int min = x[N-1];
 for(int i=1; i<N-1; i++){
  if(max < x[i]) 
   (*@\vspace{-3mm}   max = x[i]; @*)
 (*@\vspace{-3mm} \hlgray{ if(min > x[i])}   \subfloat{\usebox\mybox} @*) 
    min = x[i];	
 }
 int res = max - min;
 return res; }
\end{lstlisting}
\begin{tabular}{c}
    \textit{Test case:~} \t{largestsGap([9,5,4]) = 5}\\[1mm]
    a) Direct manipulation on 
    failing test case
\end{tabular}
\end{minipage}
\begin{minipage}{.02\textwidth}
\hspace{-7mm}
$\xrightarrow{\text{\Name}}$
\end{minipage}
\begin{minipage}{.46\textwidth}
\begin{lstlisting}[language=diff, numbersep=5pt, xleftmargin=.03\textwidth]
int largestGapFix(int[] x){ 
  int N = x.length;
  int max = x[N-1];
  int min = x[N-1];
- for(int i=1; i<N-1; i++){
+ for(int i=0; i<N-1; i++){ 
    if(max < x[i]) 
      max = x[i];
    if(min > x[i]) 
      min = x[i];	
  }
  int res = max - min;
  return res; }
\end{lstlisting}
\vspace{2pt}
\begin{tabular}{c}    
    b) Repair computed by \Name
\end{tabular}
\end{minipage}
\caption{
Examples of repair using direct manipulation in \Name.
	\label{fig:motiv}}
\vspace{-\baselineskip}
\end{figure*}

Consider the example in Fig.~\ref{fig:motiv} (a)
where a student is trying to write a program \texttt{largestGap} for finding  the \emph{largest gap} in a non-empty array of integers
---i.e., the difference between the maximum and minimum values in the array.
In the following, we assume that the student has discovered that the program behavior on test \t{[9,5,4]} is incorrect and is trying to get a suggestion from the repair tool on how she could fix the program.

\mypar{Repairing via test cases} 
Several program repair tools support test cases  as a way
to express the correct behaviour of the program.
In this case, the student can specify that on the input
\t{[9,5,4]}, the correct output should be $5$.
However, even the tool \Qlose~\cite{qlose}, which can often find correct repairs using a small number of test cases, will return  the following wrong modification to line 11:
\begin{lstlisting}[mathescape=true,numbers=none]
        int res = max - min; $\longrightarrow$ int res = max - min + 4;
\end{lstlisting}
\Qlose requires two additional test cases to find the correct repair.

\mypar{Repairing via direct manipulation} 
Direct manipulations
allow programmers to convey \emph{more information about the behavior of a test case}, rather than only its final output. 
Our technique is inspired by Guo's observation~\cite{Guo13} that 
students find it beneficial to visualize concrete program executions
and observe discrepancies between the variable values they observe and those they expect. 
For example, while debugging the \t{largestGap} program, 
the student notices that in the first iteration of the loop, right before executing line 8, variable \t{max}
has value 5 instead of the expected value 9.
While visualizing the trace, the student can directly modify the value of \t{max} as shown in the figure
and \Name will synthesize the program 
\t{largestGapFix} consistent with the manipulation (Fig.~\ref{fig:motiv}(b))---i.e., 
when running \t{largestGapFix} with input \t{[9,5,4]},  
there is a point in the execution  where the variable \t{max} contains value $9$ right before executing line 8.
Why does this new specification mechanism lead to the desired repair?
First, by modifying the program's trace and its value at line 8, the student implicitly states that certain lines
do not need modification---e.g., lines 11 and 12.
Second, the modification provides information about an intermediate state of the program that a tool cannot access  through just an input/output example.
Besides the variable \t{max}, the student can modify the value of \t{i} from 1 to 0 or the values of both \t{i} and \t{max} at the same position and \Name will produce the same repair.

Remarkably, direct manipulation can also help
debug partial implementations. Consider, for example, an incomplete version of
the program \t{largestGap} in which lines 8--9 are missing because the student has not implemented the logic for min yet.
The test case in Fig.~\ref{fig:motiv}(a) 
 is essentially useless. On the other hand, 
 the same direct manipulation shown in Fig.~\ref{fig:motiv}(a) will yield a good repair.

\section{Problem definition}
\label{sec:repair}

In this section, we define the class of programs we consider, the notion of direct manipulation, and
 our program repair problem. 

\subsection{Programs and Traces}
\label{sec:progtraces}
We consider a simple imperative language in which a program $P$ consists of a function definition 
$f(\inp_1,\ldots, \inp_q):o$
with input variables $I=\{\inp_1,\ldots, \inp_q\}$ and output variable $o$  (\t{NULL} for void functions),
a set of program variables $\varbar$ such that $V\cap I=\emptyset$, 
and a sequence of labeled statements $\sigma=s_1\ldots s_n$. 
A statement is one of the following: \verb-return-, assignment, conditional or loop statement.
Each statement in $\sigma$ is labeled with a unique 
location identifier from the set $\Loc = \{\ell_0, \ell_1, \ldots, \ell_p, \exit\}$.
We assume a universe $\mathcal{U}$ of values.
We also assume variables are associated with types and assignments are consistent with these types.

Without loss of generality, we assume that executing a \verb-return- statement assigns a value 
to the output variable 
and transfers control to a designated location $\exit$. 
A program configuration $\config$ is a pair $(\ell, \val)$ where $\ell \in \Loc$ is a location and 
$\val : \inpbar\cup\{o\}\cup \varbar\mapsto \mathcal{U} \cup \{\bot\}$
is a valuation function that assigns values to all variables. The element $\bot$ indicates that a variable 
has not been assigned a value yet or is out of scope. 
We write $(\ell, \val) \to (\ell', \val')$ 
if executing the statement at location $\ell$ 
under variable valuation $\val$
transfers control to location $\ell'$ with  
variable valuation $\val'$.
The execution trace $\exec_\Prog(\val_0)$ of the program $\Prog$ on 
an initial valuation $\val_0$ is 
a sequence of configurations $\config_0$, $\config_1$, \ldots,
where $\config_0 = (\ell_0, \val_0)$ 
and for each $h$, we have $\config_h \to \config_{h+1}$.
An execution terminates once the location $\exit$~is reached and we only consider
programs that terminate on all inputs. We use $\exec_P(\val_0)_l=\config_l$ 
to denote the configuration at index $l$  and 
$\exec_P(\val_0)_{[l,h]}$ to denote the subsequence of configurations between index $l$ and $h$---e.g.,
$\exec_P(\val_0)_{[3,5]}=\config_3\config_4\config_5$.


		Consider the program \t{largestGap} in Fig.~\ref{fig:motiv}.
	The input variable set $\inpbar$ is $\{\t{x}\}$ and the designated output variable is \t{res}.
	The set of program variables is 
	\begin{wrapfigure}{r}{0.5\textwidth} 
\begin{center}
\vspace{-11mm}
		{\footnotesize
\begin{tabular}{c|ccccccccccccc}
	& $\config_0$ & $\config_1$ & $\config_2$ & $\config_3$ & $\config_4$ & $\config_5$ & $\config_6$ & $\config_7$ & $\config_8$ & $\config_9$ & $\config_{10}$\\ \hline 
loc 		& 2    & 3  & 4 & 5 & 6 & 7 & 8 & 5 & 10 & 11 & $\exit$ \\ 
\t{N}   & $\bot$ & $3$  & 3 & 3 & 3 & 3 & 3 & 3  & 3  & 3 & 3 \\
\t{i}     & $\bot$ & $\bot$ & $\bot$ & $\bot$  & 1 & 1 & 1 & 1 & $\bot$   & $\bot$   & $\bot$  \\
\t{max}     & $\bot$ & $\bot$ & 4  & 4 & 4 & 4 & 5 & 5 & 5  & 5  & 5 \\
\t{min}     & $\bot$ & $\bot$ & $\bot$ & 4 & 4 & 4 & 4 & 4 & 4  & 4  & 4 \\
\t{res}     	& $\bot$&  $\bot$ & $\bot$  & $\bot$ & $\bot$ & $\bot$ & $\bot$ & $\bot$ &$\bot$ &$\bot$ & 1 \\ 
			\end{tabular}
			\vspace{-12mm}
		}
\end{center}	
\end{wrapfigure} 	
	$\{\t{i,max,min}\}$.
	Let $\val_0$  be the initial valuation
	such that $\val_0(x)=\{9, 5 ,4\}$ and $\val_0(w)=\bot$ for every other variable $w$.
	The execution of \t{largestGap} on $\val_0$ is illustrated on the right (we do not show valuations of the input variable $x$ 
	as $\val_h(x)=\val_0(x)$ for all $h$).

\subsection{Repair via Direct Manipulation} 

We first define the notion of direct manipulation, which allows users to express their 
 intent
by modifying 
variable values in intermediate configurations.
We assume a fixed program $P$.
A \emph{direct manipulation}
$\mathcal{M}$
 is a tuple 
$(\val_0,\ind,\val')$ where $\val_0$ is an initial valuation, $\ind$ is an index s.t. $\ind\leq |\exec_P(\val_0)|$, and 
$\val': \varbar \cup \{o\} \mapsto \mathcal{U} \cup \{\ns\}$ 
 is a new partial variable valuation.
Intuitively, the manipulation replaces
the configuration $\exec_P(\val_0)_\ind=(\ell,\val)$ at location $\ell$ with the 
the new partial configuration
$(\ell,\val')$.
Notice that a partial configuration cannot change the values of the input variables in $I$
and it can assign a special value $\ns$ to certain variables.
This value is used to denote that the manipulation ``does not care'' about the specific values of
certain variables.
We say that a valuation $\val$ \emph{satisfies} a partial valuation $\val'$, denoted $\val\vdash\val'$, iff for every variable $x\in V\cup\{o\}$, if $\val(x)\neq\ns$ then $\val(x)=\val'(x)$.

\begin{example}
\label{ex:manipulation}
The direct manipulation in Fig.~\ref{fig:motiv}(a) is formally defined as  the pair $(\val_0,6,\val')$
where $\val_0$ is the same as at the end of Section~\ref{sec:progtraces}, $\val'(max)=9$ and
$\val'(i)=\val'(min)=\val'(o)=\ns$.
This manipulation, which
modifies $\config_6$, 
only sets the value of \t{max} to 9 at location 8 and leaves all other variables unconstrained.
\end{example}

Given a program $P$, a direct manipulation $\mathcal{M}=(\val_0,\ind,\val')$ such that $\pi_P(\val_0)_\ind=(\ell,\val)$, 
we say that a program $P'$ \emph{satisfies the manipulation} $\mathcal{M}$, 
if 
there exists some index $j$ such that $\pi_{P'}(\val_0)_j=(\ell,\val_j)$ and $\val_j\vdash \val'$. In other words, a program $P'$ satisfies a direct manipulation if there exists some configuration in the execution trace of $P'$ satisfying the manipulated valuation $\val'$ at location $\ell$. 

The goal of our repair problem is to find a program that satisfies a given manipulation.
In what follows, we fix a \emph{repair model},
which is a function $\repair$ that assigns to a program a corresponding
\emph{\repspace} $\progspace$. The repair space represents a set of programs from which we can draw candidate repaired programs. 
\begin{definition}[Repair via Direct Manipulation]
\label{def:problem}
Given a program $P$ and a direct manipulation $\mathcal{M}=(\val_0,\ind,\val')$, 
the \emph{program repair via direct manipulation} problem
is to find a program $P'\in\repair(P)$ that satisfies the manipulation $\mathcal{M}$.
\end{definition}

Informally, a direct manipulation  
$(\val_0,\ind,\val')$ at location $\ell$
is a {\em reachability specification} requiring that  
a configuration $(\ell,\val')$ is eventually reached along an execution from the initial valuation $\val_0$. 
This specification mechanism is orthogonal to that provided by assertions, which
require a property $\varphi$ at location $\ell$ to be an {\em invariant specification}---i.e., each time an execution reaches location $\ell$, the property $\varphi$ should hold.  
For instance, in Figure~\ref{fig:motiv} (a), placing the assertion \t{max = 9} at location $8$ would specify
that the value of \t{max} should be $9$ at location $8$ across all loop iterations in an execution. The astute reader may suggest that for some suitably chosen predicate $condition$ over the loop counter, an assertion of the form \t{$condition \; \Rightarrow$ (max = 9)} at location $8$ could encode the direct manipulation in Figure~\ref{fig:motiv}. However, a direct manipulation does not explicitly indicate what such a predicate $condition$ should be. In particular, a direct manipulation does not specify
what the manipulation-satisfying index $j$ should be. 

\mypar{Handling test cases}
Definition~\ref{def:problem} can be generalized to the problem of repairing a program $P$ given a direct 
manipulation and a set of tests. A test $\test$ is a pair $(\val^I,\val^O)$, where $\val^I$ is a valuation over the input variables $I$ and 
$\val^O$ is a valuation over the output variable $o$ . Let $\val_0^I$ denote an initial valuation such that 
$\val_0^I(w) = \val^I(w)$ if $w \in I$ and $\bot$ otherwise. Program $P$ satisfies a test $\test$ if  
the value of the output variable $o$ at the end of an execution $\exec_P(\val_0^I)$ of $P$ on initial valuation 
$\val_0^I$ is $\val^O$, i.e., if $j = |\exec_\Prog(\val_0^I)|-1$,  $\config_j = (\ell,\val)$ and $\val(o) = \val^O$.
Program $P$ satisfies a set of tests $\T$ if it satisfies all the tests $\test \in \T$.  
The repair problem is then to find a program that satisfies both the direct manipulation and the tests.


\mypar{\Quant~Repair}
Among the many programs that satisfy a given manipulation we would like to pick the ``best''
one. To  define what it means for a candidate repaired program to be better than another,  
 we draw upon the notions of program distances proposed
in~\cite{qlose}.
We define two types of distances: syntactic and semantic distances.
Given a program $P$, a syntactic distance is a function
$f_{syn}^P:\progspace\to\naturalnumber$ that maps each program
in the repair space to a quantity capturing 
its syntactic similarity to the original program $P$.
We define semantic distances using distance functions over execution traces.  
Let $\trdist(\tr,\tr')$ denote a distance function mapping a pair of traces to a non-negative integer. Intuitively, $\trdist$
captures the similarity between execution traces of $P$ and $P'$ on the same initial valuation $\val_0$. 
Given a program $P$ and a direct manipulation $\mathcal{M}= (\val_0,\ind,\val')$, 
a semantic distance function $f_{sem}^{P,\mathcal{M}}:\progspace\to\naturalnumber$ maps a repaired program $P'$  
to $\trdist(\pi_P(\val_0)_{[0,\ind]},\pi_{P'}(\val_0)_{[0,j]})$ 
capturing the similarity between the manipulated trace $\pi_{P}(\val_0)_{[0,\ind]}$ of $P$ 
and the corresponding {\em manipulation-satisfying trace} $\pi_{P'}(\val_0)_{[0,j]}$ of $P'$ with manipulation-satisfying index $j$. 
An aggregation function $\textsc{Aggr}: \naturalnumber\times \naturalnumber\rightarrow\naturalnumber$
is used to combine the two distance functions. 
 
\begin{example}
\label{ex:distances}
An example of syntactic distance between two programs $P$ and $P'$ is the number of node edits needed to transform
 the abstract syntax tree $P$ into the one $P'$. 
 According to this distance, the change from \t{i=1} to \t{i=0} showed in 
 Figure~\ref{fig:motiv} has syntactic distance $1$. An example semantic distance is the sum of the differences in variable valuations in program configurations of the execution traces $\pi_{P}(\val_0)_{[0,\ind]}$ and $\pi_{P'}(\val_0)_{[0,j]}$ (with $j$ as defined above). 
\end{example}

\noindent For a program $P$ and direct manipulation $\mathcal{M}$, 
we can define the {\em cost} of a repaired program $P'$ as 
$cost(P')=\textsc{Aggr}(f_{syn}^{P}(P'),f_{sem}^{P,\mathcal{M}}(P'))$.
The following definition can be generalized to incorporate a set of tests.

\begin{definition}[\Quant~Repair via Direct Manipulation]
\label{def:prob}
Given a program $P$ and a direct manipulation $\mathcal{M}$, 
the \emph{\quant~program repair via direct manipulation} problem
is to find a program $P'\in\repair(P)$ that satisfies the manipulation $\mathcal{M}$
and such that,
for every $P''\in\repair(P)$ that satisfies the manipulation $\mathcal{M}$,
we have $cost(P')\leq cost(P'')$.
\end{definition}



\section{\Name's architecture}
\label{sec:architecture}

\begin{figure*}[!t]
\includegraphics[width=\linewidth]{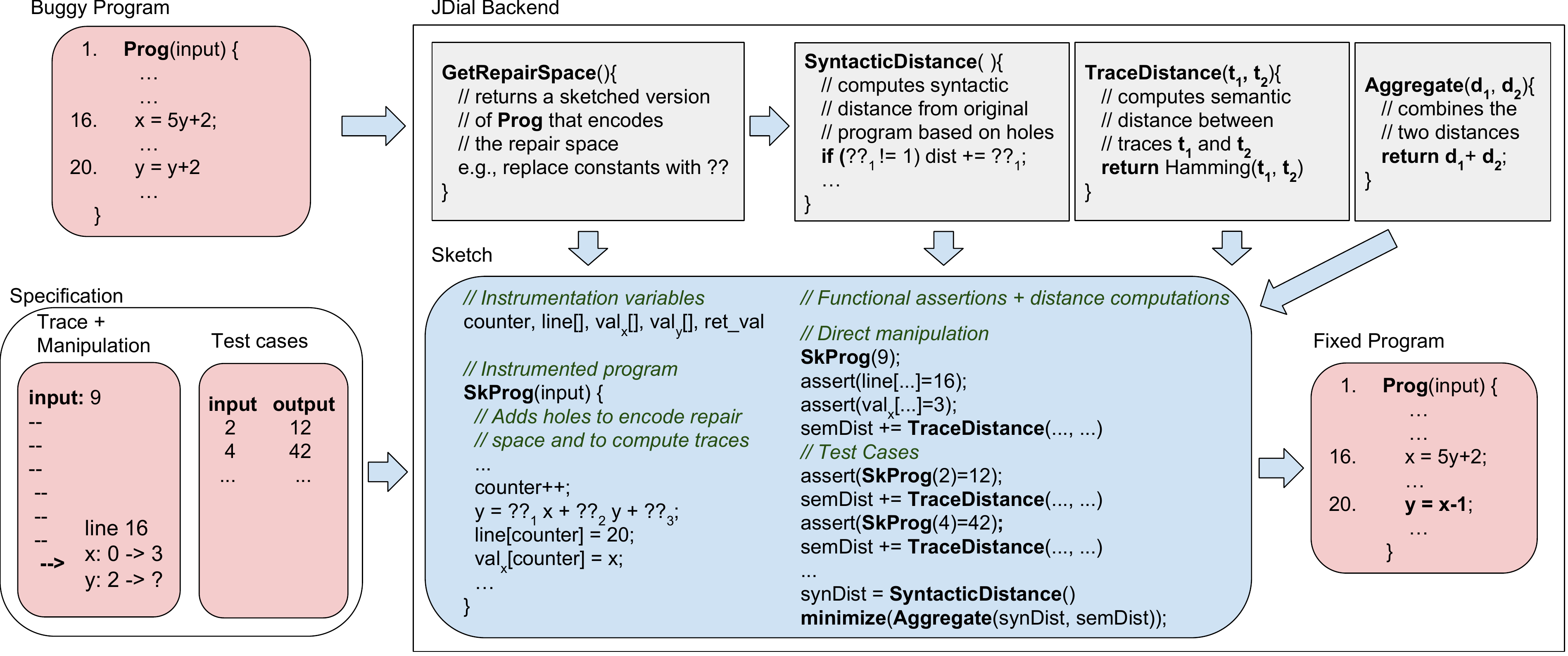}
\caption{Architecture of \Name.
 \label{fig:architecture} }
\end{figure*}

In this section, we describe the architecture of \Name
and the sketching-based approach
 \Name employs to repair programs (Figure~\ref{fig:architecture}).

\Name takes as input a
buggy program, a direct manipulation on an input trace, and (optionally)
a set of test cases (left of Figure~\ref{fig:architecture}).
As described in Section~\ref{sec:repair},
the repair problem is defined using four components:
a repair model, 
a syntactic distance function,
a semantic distance function,
and 
a cost-aggregation function.
In \Name, these components are modular and 
defined independently from the underlying synthesis engine
(grey boxes at the top of Figure~\ref{fig:architecture}).
The repair model is given as a program GetRepairSpace that, given a program, returns a sketched version of it---i.e.,
a program with unknown holes of the form \t{??}. In Figure~\ref{fig:architecture}, this program simply replaces each constant with a hole.
By  instantiating the holes in the sketched program with concrete values we obtain a program in the repair space.
The syntactic distance is given as a program that
computes a non-negative integer based on the values of the holes in the sketched
program---e.g., how many constants were changed or by how much they were changed.
The semantic distance is given as a program that computes a non-negative integer 
based on the value of two traces---e.g., the Hamming distance. 

\subsection{From Program Repair to Sketching}

To solve the repair problem,
\Name computes a sketched program together with a set of assertions (blue box in Figure~\ref{fig:architecture}).
The solution to this sketched program---i.e., values for the holes that 
satisfy the assertions and minimize the given objective function---is the solution to the
program repair problem.

\mypar{Background on sketching}
Program sketching
is a technique for specifying a parametric set of programs. 
This is done by allowing programs to contain 
 holes (denoted by \t{??}).
When one provides a specification---e.g., test cases, assertions, minimization objectives---the
sketching problem is to find (typically integer) values of the holes that satisfy the given specification.
State-of-the-art sketching tools support complex program constructs, such
as arrays, strings, and recursive functions, as well as complex specification mechanisms,
such as Boolean assertions and quantitative 
optimization constraints over the values of the holes~\cite{sketch}.


\mypar{Computing distances and guessing trace lengths}
The GetRepairSpace component, given the buggy program, adds holes 
to generate a sketched program encoding the repair space---e.g.,  \t{y = ??\_1 * x + ??\_2 * y + ??\_3} in Figure~\ref{fig:architecture} (blue-box).
\Name then generates a function
that uses the values placed in the holes to compute the syntactic distance  (Figure~\ref{fig:architecture}(top)). 

To compute 
the semantic distance
\Name symbolically extracts program traces by instrumenting the sketched program with 
a counter to measure the length of the trace, 
an array to record the values of each variable in the original program, and an array for the line numbers.\footnote{
We assume that the length of the trace in the repaired program is 
at most twice the length of the original trace and we use this assumption to initialize
the length of the arrays. This constant is parametric and can be modified.}
After executing each sketched instruction, the arrays are updated to reflect the current variable values (Figure~\ref{fig:architecture}(blue-box)).
\Name then extract the traces from such arrays and uses them to compute
the semantic distance.


The key difficulty in encoding our repair problem  is that there can be many ways to ``align'' the location manipulated by the user
with a location in the sketched program---e.g., the execution of one repaired program might reach the desired manipulated value 
the second time
the manipulated location  is visited, while another repair might
reach the desired value the tenth time the manipulated location is visited.
\Name must be able to consider all these possibilities.

\begin{example}
Consider the manipulation described in Fig.~\ref{fig:motiv}(a). 
The execution of the repaired program \t{largestGapFix} presented in Fig.~\ref{fig:motiv}(b) on 
input $[9,5,4]$
hits the manipulated location with \t{max=9} the first time line 8 is traversed.
However, another correct repair which changes the loop in line $5$ 
to \t{i=N-2; i>=0; i-\hspace{0mm}-}, hits the manipulated location with \t{max=9} the second time line 8 is traversed.
\end{example}

\begin{wrapfigure}{r}{0.29\linewidth} 
\vspace{-10mm}
\lstset{xleftmargin=2pt}
\begin{lstlisting}
...
if(visit_time == 0) {
   // record state 
   return; 
}
else 
   visit_time--;
//Sketched version of 
//manipulated line
...
\end{lstlisting}
\vspace{-2mm}
\caption{Instrumentation to guess  visiting times.
\label{fig:visiting}}
\vspace{-10mm}
\end{wrapfigure}
Our key idea is to introduce an existential unknown variable---i.e., another hole---in our sketched program to guess at what visit time the
manipulated line is reached with the variable values provided by the user.
Concretely, we define a global variable $\t{int visit\_time = ??}$ to guess the number of visits of the manipulated line and modify  the sketched program
right 
before the sketched version of the manipulated line to interrupt the trace at
the correct time (see Fig.~\ref{fig:visiting}). Thus, every time the manipulated
line is reached, \t{visit\_time} is decremented and, when the counter hits zero, the execution
has reached the guessed number of visit times.

Finally, \Name adds assertions to guarantee the sketch solution
satisfies the manipulation and a minimization objective to ensure the returned 
solution is optimal with respect to the given distances
(see right of blue box in Fig.~\ref{fig:architecture}).

\section{Implementation and optimizations}
\label{sec:implementation}

%
\Name is composed of a frontend, which
allows to visualize program traces and manipulate intermediate states,
and a backend, which repairs the manipulated programs.
\Name's frontend uses 
the tool \textsc{java\_jail} to extract program traces and
JavaScript, together with the library CodeMirror, to render the user interface and communicate with the backend.
\Name's backend
uses the synthesizer \textsc{Sketch}~\cite{sketch} to solve the sketched programs we generate
using the technique proposed in Section~\ref{sec:architecture}.
\Name can repair
Java programs over integers, characters, Booleans, and arrays over these basic types.
In its default mode, \Name only tries to repair statements in the function in which the manipulated line appears. 
In this section, we describe
 the concrete repair model and distance functions \Name uses as well
 as several optimizations employed by \Name.


\subsection{Repair Model and Syntactic Distance}
\Name  supports complex repair models---e.g., it can allow
statements to be added to the program.
However, 
overly expressive repair models will often lead to undesired
repairs that overfit to the given manipulation.
In fact, existing repair tools for introductory programming
assignments typically employ several repair models, each tailored to a particular programming assignment~\cite{SGS13,rolim2016learning}.

Since in our application domain we do not know a priori what program the programmer
 is trying to write,
 \Name's default repair model only allows to rewrite constants in linear arithmetic expressions (see Appendix~\ref{sec:repair-model} for details). 
First, any 
variable in any expression is multiplied by a hole \t{??$_b$} that can only take values from the set 
$\{-1,0,1\}$. 
These holes can be used to remove variables and negate their coefficients.
Second, the term 
\t{$\sum_{\texttt{v}\in V}$??$_{b}$v+??}, where $V$ is the set of variables, is added to
each expression appearing in an assignment or in a Boolean comparison.
These terms can be used to add new variables, further increase/decrease the coefficients
of variables appearing in the expression, or add new constants---e.g., turn \t{x<0} into \t{x>y}. 
This repair model permits modifications of multiple expressions and it subsumes the default error model
of the \textsc{AutoGrader} tool, which, despite its simplicity,
was shown to be able to repair 30\%-60\% of edX student submissions depending on the problem type~\cite{SGS13}.

\Name's syntactic distance computes the difference between the synthesized hole values  and the original ones.
For example, in the expression
\t{??$_b$x < 0 + $\sum_{\texttt{v}\in V}$??$_{b}$v+??} (corresponding to original expression \t{x<0}), 
the original value of the first hole $??_b$ is 1, while the original value of all the other holes is 0.
The syntactic distance is the sum of the absolute difference
between each hole's synthesized value and the original one.
Intuitively, this distance penalizes modifications that introduce new variables and modify constants by large amounts.

%

\subsection{Semantic Distance over Traces}
\label{subsec:repairinjdial}

When computing the distance from the original program traces, $\Name$ ignores the variables that have been manipulated
because they are likely to contain ``incorrect'' values that are not necessary
to preserve. 
\Name first computes the restricted traces where the values of the manipulated variables
are omitted and then uses a modified version of the Hamming 
distance to compute their distance. 
In the following definitions, we assume Boolean tests return 1 when true and 0 when false.
Given two configurations $\config=(l,\val)$ and $\config'=(l', \val')$ over a set of variables $V$,
the distance  between the two configurations is defined 
as $H(\config, \config') = (l\neq l') + \sum_{w\in V}\val(w)\neq \val'(w)$.
Finally,
\Name computes the distance between two traces $\pi=\config_1\cdots \config_s$ and 
$\pi'=\config_1'\cdots \config_t'$, where
 $m=min(s,t)$ and $M=max(s,t)$  as the quantity
$H(\config_1,\config_1')+\cdots +H(\config_m,\config_m')+M-m$.

\begin{example}
\label{ex:costfunction}
Consider again the example described in Figure~\ref{fig:motiv}. 
The restricted trace of the repaired program up to the manipulation-satisfying index 
has distance 3 from the original trace since it only changes the value of  the variable \t{i} in the last three
steps and it has the same length as the original program trace.
%
\end{example}

\Name contains other implementations of trace distances---e.g., longest common subsequences. 
Since the distance presented above yields good results and performance in practice, we use it as default and 
use it in our experiments.
\Name aggregates the syntactic and semantic
 distances by taking their sum. 

\subsection{Handling External Functions}
\label{subsec:auxiliary}
\Name  employs a new 
Counterexample-Guided Inductive Synthesis (\textsc{CeGIS}) scheme to 
 repair programs that contain external functions for which semantics might be unknown or expensive to encode directly in \textsc{Sketch}. 
Given the input program with an external function \t{ext} and the manipulated trace, \Name creates a sketched program that 
 assigns a partial interpretation to the external function using
 the set of concrete values obtained from the input trace execution---i.e., for every call of the function observed in the input trace. 
\Name then computes a repair for the sketched program using the partial definition of \t{ext} it obtained from
the trace. If repairing the program requires knowing the interpretation of \t{ext} on inputs that have not been observed
yet, \Name lets \sketch ``guess'' an interpretation for the function \t{ext}  on such inputs.
\Name can then execute the function \t{ext} and
check whether the guesses were correct.
If they are not correct,
\Name modifies the new sketched program to incorporate the partial interpretation to the external function \t{ext}
on the newly discovered inputs. The process continues until \Name finds a repair that respects the semantics of \t{ext}.

\begin{figure*}[!t]
\centering
	\begin{tabular}{cccc}
r			\begin{lstlisting}
// Test case
// sumPow(3)=15
int sumPow(int x){
  int sum = 1;
  for(int i=1;i<x;i++){
    sum+=Math.pow(2,i);
  }
  return sum;
}
\end{lstlisting}
		&		
				\begin{lstlisting}[mathescape=true]
// Partial interpretation 
// of Mathpow
int Mathpow(a, b){
  if(a==2 && b==1)
    return 2;
  if(a==2 && b==2)
    return 4;
  if(a==?? && b==??)
    return ??; }
				\end{lstlisting}
&
{
\scriptsize	
		\begin{tabular}{l}
		1. \Name guesses \t{Mathpow(2,3)=14}\\
		2. Is \t{Math.pow(2,3)=14}?\\
		3. No, modify \t{Mathpow }\\
		\end{tabular}
		}

				\\
			(a) & (b) & (c) \\
		
	\end{tabular}
	\caption{
	Given an example with an incorrect for-condition and an input test (a), \Name uses the execution of \t{sumPow} on the test
	 to learn
	 an initial partial interpretation of the
	function  $\t{Math.pow}$ (b). 
\Name then produces a proposed repair and  guesses the interpretation of \t{Math.pow}
to be such that \t{Math.pow(2,3)=14} (c).
After executes \t{Math.pow} in Java, \Name discovers that \t{Math.pow(2,3)=8}, and refines the interpretation of $\t{Math.pow}$ for the next round
of repair.}
	\label{fig:cegisexample}
\end{figure*}

\begin{example}
Consider the program \t{sumPow} in Fig.~\ref{fig:cegisexample}(a),
which should compute the sum of powers
of 2
up to \t{x}, but
instead it only computes the sum up to \t{x-1}.
By running the program on the given input 3, \Name can obtain the output of the $\t{Math.pow}$ function on input values 1 and 2, and constructs a \sketch function that describes a partial interpretation of \t{Math.pow} as shown in Figure~\ref{fig:cegisexample}(b).
To repair the program, \Name needs to change the condition of the for loop,
but this repair requires knowing the output of the function
$\t{Math.pow}$ on arguments (2,3) and our partial interpretation of $\t{Math.pow}$ does not contain this information. 
\Name synthesizes a repair for the function \t{sumPow} and, while doing so, it assigns an interpretation
to the inputs for which the behaviour of the function \t{Math.pow} is unknown (Figure~\ref{fig:cegisexample}(c)). 
\Name then uses  the concrete execution of the function \t{Math.pow}
to check whether the synthesized interpretation is incorrect, and in this case it modifies the partial
interpretation of \t{Math.pow} in the sketched program. 
\end{example}

\subsection{Additional Features and Optimizations}
\label{subsec:optimizations}

We discuss  additional features and optimizations  supported by \Name . 

\mypar{Specified repair range} 
Since the programmer might want to prevent \Name from modifying certain 
program statements, \Name's frontend allows the programmer to specify what
statements 
the tool is allowed to modify.

\begin{wrapfigure}{r}{0.52\linewidth} 
	\lstset
	{ 
		numbers=left,
		stepnumber=1,
		tabsize=2,
	}
	\vspace{-9mm}
\begin{lstlisting}[language=diff,escapeinside={(*@}{@*)}]		
// input: [3,2,7]
int[] subLargestGap(int[] a){
 int N = a.length;
 int min = max = a[0];
 for(int i=0; i < N; i++){
  if(max < a[i]) max = a[i];
  if(min > a[i]) min = a[i];}
 int largestgap = max - min;
(*@R:\hlgray{for(int i=1; i<N; i++)\{ }@*)     
   a[i] = a[i]-largestgap;
(*@M:\hlgray{\}  }  @*)     
 return a; }
\end{lstlisting}
\vspace{-2mm}
\begin{lstlisting}[language=diff]		
void sliced(int[] a){
 int N = 3;
 int largestgap = (*@\textbf{5};@*) 
 for(int i = 1; i < N; i++)
  a[i] = a[i]-largestgap;
 return a; }
\end{lstlisting}
	\caption{Program \t{subLargestGap} and
	its \t{sliced} version 
	when the manipulation happens at line 13 and only line 11 can be modified.}
	\label{fig:slicing-example}
\end{wrapfigure}
\mypar{Single statement repairs}
Since most repairs only require
to  modify a single statement, \Name supports this restricted repair
model and it uses an optimized solver that,
 for each line of code, builds a separate sketched program 
that is only allowed to modify that line.
The separate sketched programs are then solved in parallel and
 \Name outputs the repair
with the least aggregate cost.

For each individual sketch that can only modify a certain line of code, 
 \Name uses a data flow analysis based on program slicing~\cite{Weiser81}  to 
summarize parts of the program for which the corresponding traces will not be affected by the code modification.
Concretely, let $\ell_{\mathcal{M}}$ be the location at which the manipulation is performed and $\ell_R$ be the location
 \Name is allowed to modify.
By computing a \emph{backward} slice of the manipulated  location
$\ell_{\mathcal{M}}$, we obtain the statements that can affect the values of the manipulated variables.
Similarly, only statements that are reachable from location $\ell_R$ in the control-flow 
graph of the program  are affected
by modifications to line $\ell_R$. 
Finally, the intersection of the two sets gives us the 
statements where variable values may vary as a result of a repair. All other statements are irrelevant and can be removed or summarized.

\mypar{Example 6}
Consider the program \t{subLargestGap} in Fig.~\ref{fig:slicing-example}
that
returns a new array obtained by  subtracting the largest gap of the input array from all its elements.
This program contains a mistake in the second for loop.
Assume a student is trying to fix it by manipulating the variable \t{a[0]} at location 11
on input \t{[3,2,7]}
and that the repair model only allows modification to location 9.
The backward slice of location 11 contains all the statements in the program except the return statement
and
the lines 9 to 12 are the only lines reachable from location 9.
Using this information, \Name summarizes all other statements' values. For example, the whole computation of
the variable \t{largestGap} is replaced by the constant assignment \t{largestGap=5}.

\lstset{language=[Sharp]C,
  showspaces=false,
  showtabs=false,
 breaklines=true,
  showstringspaces=false,
 breakatwhitespace=true,
  escapeinside={(*@}{@*)},
  commentstyle=\color{greencomments},
  keywordstyle=\color{bluekeywords},
  stringstyle=\color{redstrings},
  basicstyle=\scriptsize\ttfamily,
  morekeywords={str,ch,bit,harness, assert,loc},
  xleftmargin=5ex
}

\section{Evaluation}
\label{sec:evaluation}

We evaluate the effectiveness of \Name through the following questions. 

\vspace{2mm}
\hspace{-5mm}
\begin{tabular}{p{0.05\linewidth}p{0.95\linewidth}}
\textbf{Q1} &  Can \Name produce good repairs more often than test-based techniques? 
\\

\textbf{Q2} & Is the optimized version of
\Name
presented in Sec.~\ref{subsec:optimizations} effective? 
\\

\textbf{Q3} &  How  sensitive is \Name w.r.t. the location of the
 manipulation?
\\

\textbf{Q4} & Can \Name 
				repair programs that contain external functions?
\end{tabular}
\vspace{1mm}

\setlength{\tabcolsep}{2pt}
\begin{table}[!t]
\caption{
Effectiveness and performance of \Name.
\textbf{Success} denotes the number of tests (out of 5) for which we get a desired repair.
\xmark denotes out of memory.
\comm{
\Name. We report numbers of lines of code (LOC), program variables (Vars), 
test cases out of 5 for which we get a desired repair (Success), length of manipulated trace,
average running time.
The last two columns show the running times for unoptimized (\Namesingle) and optimized (\Namesingleopt) versions
of \Name when the repair model allows to modify at most one statement. 
\xmark denotes out of memory.
}}
\footnotesize
\centering
\begin{tabular}{cc|ccccr|rr}
& \multirow{2}{*}{\bf Problem} & \multirow{2}{*}{\bf LOC} & \multirow{2}{*}{\bf Vars} & \multirow{2}{*}{\bf Success} &{\bf Trace} & \multicolumn{1}{C{10mm}}{\bf Time} &  \multicolumn{2}{|c}{\bf Time single line} [sec]\\
&    &    &    &    & \textbf{Length}    &\multicolumn{1}{c}{ [sec]}    & \multicolumn{1}{|c} ~~\Namesingle &\multicolumn{1}{c}{\Namesingleopt}\\
\hline 
\parbox[t]{2mm}{\multirow{12}{*}{\rotatebox[origin=c]{90}{\Qlose~\cite{qlose}}}} 
&\t{largestGap-1.1} & 7 & 4      &  3     &   11    & 3.8\hspace{10pt}  &1.6\hspace{2mm}        & 1.0\hspace{5mm}\ \\ 
&\t{largestGap-1.2} & 7 & 4      &  4     &   10    & 2.2\hspace{10pt}  &0.8\hspace{2mm}         & 0.6\hspace{5mm}\ \\ 
~~~~~&\t{largestGap-2} & 7 & 4     &  2     &      15    & 4.2\hspace{10pt}   &1.1\hspace{2mm}     & 0.5\hspace{5mm}\ \\ 
&\t{largestGap-3.1} & 7 & 4     &  4     &      10    & 1.8\hspace{10pt} &1.1\hspace{2mm}        & 0.5\hspace{5mm}\ \\
&\t{largestGap-3.2} & 7 & 4     &  4     &      10    & 2.8\hspace{10pt}  &1.0\hspace{2mm}        & 0.6\hspace{5mm}\ \\
&\t{tcas} & 10 & 4     &   0    &     7    & 0.8\hspace{10pt}          &0.4\hspace{2mm}          & 0.4\hspace{5mm}\ \\
&\t{max3} & 5 & 3  & 5                  &      3     & 0.5\hspace{10pt}     &0.3\hspace{2mm}    & 0.3\hspace{5mm}\ \\ 
&\t{iterPower-1} & 5 & 3     &  4     &      14    & 0.4\hspace{10pt}          &0.6\hspace{2mm}        & 0.4\hspace{5mm}\ \\ 
&\t{iterPower-2} & 5 & 3     &  5     &      14    & 0.7\hspace{10pt}       &0.4\hspace{2mm}    & 0.3\hspace{5mm}\ \\ 
&\t{ePoly-1} & 6 & 4          &  4     &   12    & 4.6\hspace{10pt}           & 3.7\hspace{2mm}          & 1.3\hspace{5mm}\ \\ 
&\t{ePoly-2} & 6 & 4         &  3      &     12    & 2.5\hspace{10pt}          & 1.7\hspace{2mm}        & 0.9\hspace{5mm}\ \\ 
&\t{multIA} & 4 & 4         &  5     &   9     & 1.3\hspace{10pt}         &0.8\hspace{2mm}     & 1.1\hspace{5mm}\       \\ 
\cdashline{1-9}
\parbox[t]{2mm}{\multirow{4}{*}{\rotatebox[origin=c]{90}{\!\!\!\!\!\!New}}}
&\t{ePoly-3} & 7 & 4         &  4        &     13    & 2.9\hspace{10pt}     &2.8\hspace{2mm}       & 2.5\hspace{5mm}\  \\ 
&\t{max4} & 7 & 4             &  5     &     4    & 0.3\hspace{10pt}         &0.2\hspace{2mm}         & 0.3\hspace{5mm}\   \\ 
&\t{bubbleSort} & 7 & 5      &  4     &     12    & 3.1\hspace{10pt}         &1.3\hspace{2mm}               & 0.6\hspace{5mm}\      \\ 
&\t{subLargestGap} & 13 & 6      &  \xmark     & 35    & \xmark\hspace{10pt}  &\xmark\hspace{2mm}               & 0.7\hspace{5mm}\      \\ 
&\t{maxMin} & 13 & 6      &  \xmark     & 37    & \xmark\hspace{10pt}        &\xmark\hspace{2mm}           & 0.9\hspace{5mm}\      \\ 
\end{tabular}
\label{fig:results}
\vspace{-2mm}
\end{table}

We perform our evaluation on 17
Java programs: 12 from \Qlose~\cite{qlose} and 
5 new programs. 
The programs from~\cite{qlose} are small introductory programming assignments and 
programs from repair literature (see Appendix~\ref{sec:benchmark} for details). 
  
Table~\ref{fig:results} shows detailed metrics for each benchmark and the average runtime of \Name when performing repair 
on five randomly generated failing inputs. 
All experiments were performed on an Intel Core i7 4.00GHz CPU with 32GB/RAM.

\bpar{Comparison to Test-based Repair}
We compare \Name against the tool \Qlose to
see if repair via direct manipulation can find correct repairs more often than test-based repair.
We compare against \Qlose because  it is the only repair tool that uses semantic distances
and it produces accurate repairs from a small number of test cases more often than tools that only
use syntactic distances~\cite{qlose}.

For each benchmark, we randomly generate 5  input tests that result in incorrect outputs.
For each failing test,
we run \Qlose using the test as a specification
and run \Name
by manually constructing a manipulation: we identify the first location in the execution trace 
where a variable has the wrong value and modify it to the correct one. 
Figure~\ref{fig:comparedWithQlose}  illustrates the results of this comparison (\Name and \Qlose bars).
\Name generates the intended repairs in $66\%$ (56/85) of the cases while
\Qlose never produces a correct repair. In particular, when given only one test case, \Qlose always modifies the return statement of the program.

\begin{figure}[!t]
\scriptsize
\begin{tikzpicture}
\begin{axis}[
ybar=0pt,
bar width=.8mm,
width=\textwidth,
height=.28\textwidth,
legend style={at={(0.5,1.15)},
	anchor=north,legend columns=-1},
symbolic x coords={largestGap-1.1,largestGap-1.2,largestGap-2,largestGap-3.1,largestGap-3.2,tcas,max3,iterPower-1,iterPower-2,ePoly-1,ePoly-2,multIA,ePoly-3,max4,bubbleSort, subLargestGap, maxMin},
xtick=data,
ymin=0,ymax=6,
ylabel={Succesful tests},
x=16pt,
xticklabel style = {font=\ttfamily,rotate=40,anchor=east},
]
\addplot table[x=interval,y=Qlose]{endVSmiddle.tab};
\addplot table[x=interval,y=JDial]{endVSmiddle.tab};
\addplot table[x=interval,y=Qlose+]{endVSmiddle.tab};
\addplot table[x=interval,y=JDial+]{endVSmiddle.tab};
\legend{ \Qlose, \Name, \Qlose\!\!+, \Name\!\!+}
\end{axis}
\end{tikzpicture}
	\vspace{-3mm}
	\caption{
	\Name vs \Qlose. Additional test provided in \Name\hspace{-1mm}+ and \Qlose\hspace{-1mm}+.
	}
	\label{fig:comparedWithQlose}
	\vspace{-1.2\baselineskip}
\end{figure}

We perform another study where, for 
each previous experiment,
 we provide \Name and \Qlose with an additional (failing or passing)
test---i.e., we provide \Qlose with two tests and
\Name with one test and one manipulation.
Figure~\ref{fig:comparedWithQlose}  illustrates the results of this comparison (cf. \Namep and \Qlosep bars).
\Name generates the intended repairs for $75\%$ (64/85) of the cases while
\Qlose produces the correct repairs on $58\%$ (49/85) of the cases.
While \Qlose performs better 
than when given a single test, 
for every input on which \Qlose produces the correct repair, \Name also does so.
Remarkably, when given a single manipulation and nothing more, \Name produces correct repairs 
more often than \Qlose, even when the latter is provided with 2 tests.
To answer \textbf{Q1},
\textbf{JDial produces meaningful repairs more often than techniques that only use tests}.

Before concluding,   we explain why both tools performed poorly on some benchmarks.
For the  \t{tcas} program, the desired fix modifies an expression by adding a
large constant that can only be synthesized from a very specific test case. 
\t{subLargestGap} and \t{maxMin} are too large.
For the instances for which \Name produces the incorrect repair,
we evaluate whether \Name  produces correct repairs if it is allowed 
further ``attempts''. 
Whenever an undesired repair is generated at a location $\ell$, we 
disallow \Name to repair location $\ell$ again or reject the
proposed repair and ask for a different one. 
This approach repairs 
an additional 6 failing tests across all programs with an average of $2.2$ user interactions.

\bpar{Optimizations for Single-line Repairs}
We repeat the previous experiment using the single-line repair model described in
Sec.~\ref{subsec:optimizations}. We refer to the version of \Name 
with this restricted repair model as \Namesingle
and its optimized version as \Namesingleopt.
Since in our previous experiments 
\Name always finds a repair that only involves a single line, both \Namesingle and
\Namesingleopt find the same repair. 
The last two columns of Table~\ref{fig:results} show the running times.
\Namesingle is generally faster than the version of \Name
that uses the more complex repair model.
However, the optimized version \Namesingleopt
is on average 1.37x faster than \Namesingle. 
Moreover, for
  \t{subLargestGap} and \t{maxMin},
  \Namesingleopt finds repairs in <1 second while \Namesingle times out.
The improvement is due to the
slicing-based data-flow analysis, which, can  reduce the number of lines in the sketched program from 25 to 8.

To answer \textbf{Q2},
\textbf{the optimization from Sec.~\ref{subsec:optimizations} is beneficial} for single-line repairs.
This repair model is very practical and our results hint that our slicing technique
 can make \Name scale to larger programs.

\bpar{Sensitivity of Manipulated Location}
One of the key aspects of \Name 
is that the user has to find a ``good'' location to perform the desired repair.
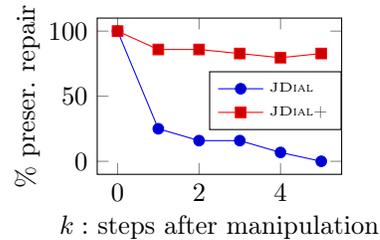
\begin{wrapfigure}{r}{.44\linewidth}
\vspace{-8mm}
	\centering
	\begin{tikzpicture}
	\begin{axis}[
	ylabel absolute, ylabel style={yshift=-3mm},
	xlabel absolute, xlabel style={yshift=1mm},
	width=.9\linewidth,
	height=.3\textwidth,
	xlabel=$k:$ steps after manipulation,
	ylabel=\% preser. repair,legend cell align={left}, legend style={font=\tiny,at={(1,0.66)}}]
	\addplot table [mark=triangle,x=step, y=rate]{differentStop.tab};
	\addplot table [mark=*,x=step, y=rate]{differentStopPlus.tab};
	
	\addlegendentry{\Name}
	\addlegendentry{\Namep}
	\end{axis}
	\end{tikzpicture}
	\vspace{-2mm}
	\caption{Correct repair if
	manipulation is performed $k$ steps later.
	}	
	\label{fig:rateDifferentPosition}
	\vspace{-9mm}
\end{wrapfigure} 
In this experiment, we evaluate how sensitive $\Name$ is with respect to the location choice. 
We consider the experiment we performed against test-based repair
and for each test  case on which \Name and \Namep successfully
 found a correct repair, we then perform the following analysis:
if the manipulation was performed at  step $i$
 in the 
program trace, we measure after how many steps the generated repair is ``lost''---i.e.,
we compute the smallest 
$k$ for which performing the manipulation at position
$i+k$ would yield a wrong repair.

Figure~\ref{fig:rateDifferentPosition} shows the results.
In 80\% of the cases, if \Name is provided only with a manipulation
and the manipulation is performed one step later, 
\Name returns an incorrect repair.
However, when provided with one additional test case
\Name returns the correct repair in 80\% of the cases,
even when the manipulation is performed 5 steps after the ideal location.
Even in these extreme conditions, \Name returns correct repairs more often
than \Qlose does when provided with two test cases.
To answer \textbf{Q3},
\textbf{\Name is sensitive with respect to the manipulation location only if no additional
tests are provided}, but it is still more precise than repair via test cases.

\bpar{Ability to Handle External Functions}
We evaluate if \Name can repair programs
with external functions.
\t{ePoly-1} and \t{ePoly-2}, contain the function \t{Math.pow}
and \Name is able to produce a repair for them using between 2 and 5 iterations (average 4.2), 
 of the \textsc{CeGIS} algorithm presented in Sec.~\ref{subsec:auxiliary}.

To better evaluate the algorithm, we design two families of benchmarks that use auxiliary functions.
The first family of programs tries to compute $\sum_{i=0}^n \t{Math.pow}(2,i)$ 
for values of $n$ between 2 and 8. The bug in this benchmark is the one shown in 
Figure~\ref{fig:cegisexample}. 
For inputs 2 and 3, \Name can find the correct repair that is compliant with the external function
after 2 \textsc{CeGIS} iterations, while
for inputs 4 through 8, \Name requires 3 iterations.
The second family of programs computes the maximum value in an array using the \t{Math.max} function  for different incorrect initializations of the variable \t{max}.
In this case, 
the size of the initial constant affects the number of required \textsc{CeGIS} iterations.
While incorrectly initializing \t{max} to 2 only requires a couple of iterations to produce the correct repair,
if we incorrectly initialize \t{max} to 100,
computing the repair requires guessing many new interpretations
of the function \t{Math.max} that did not appear in the original trace, resulting in more than 90  iterations.

To answer \textbf{Q4},
\textbf{\Name can repair programs that contain external functions}, but in certain pathological
cases it requires many \textsc{CeGIS} iterations.

\section{Related Work}
\label{sec:related}

\mypar{Direct manipulation}
Direct manipulation is an old concept~\cite{Shneiderman83} and several
drawing editors~\cite{victor,schachman,Hottelier14}
 provide programming capabilities and at the same time allow the user to interact with the graphics being displayed by the code. \textsc{Sketch-n-Sketch}~\cite{Chugh16,Hempel16}
uses a simple form of program synthesis to
combine direct and programmatic manipulation of scalable vector graphics (SVG), where constants in the program can be modified conforming to the direct manipulations. \textsc{Sketch-n-Sketch} and \Name are somewhat incomparable due to the different domains, but we describe some high-level differences between the two tools.
\textsc{Sketch-n-Sketch} can only rename constants defined at the top of the program 
and, unlike \Name, cannot handle complex updates involving changes in the program structure---e.g.,
replacing $x=y$ with $x=y-z$.
Moreover, \textsc{Sketch-n-Sketch} employs heuristics to identify the ``right'' fix, while 
\Name identifies fixes that are optimal with respect to syntactic and semantic
distances.

Wolverine~\cite{Verma17} allows direct manipulation of data-structure transformations.
In Wolverine, the user can modify a graphical abstract representation of a data structure such as a linked list and the tool
will attempt to find a program modification consistent with the modification. 
Similar to \textsc{Sketch-n-Sketch}, Wolverine's technique is specific to certain families of data structure transformations and relies on the graphical abstraction used for the manipulation.

\textsc{CodeHint}~\cite{Galenson14} synthesizes simple Java expressions---e.g., library calls---at user-set breakpoints using partial specifications---e.g., variable types. It uses  information from the execution to construct expressions of a user-provided type. \textsc{CodeHint} is different from \Name in three main aspects:
\rone \textsc{CodeHint}  helps programmers auto-complete  function calls given some expected type at a given location, whereas \Name repairs the original program using a global analysis.
\rtwo \textsc{CodeHint}
can only synthesize one expression at a time due to its specific interaction mode.
\rthree \textsc{CodeHint} performs brute-force search while \Name uses constraint-based search with optimization objectives.

\mypar{Personalized education}
There are many tools for teaching programming 
that help with grading (see~\cite{striewe2014review} for a survey), 
personalized feedback~\cite{RK15,GRZ16clara,Kim16,SGS13,qlose,rolim2016learning}, and visualization~\cite{Guo13}. 
Several works have dealt with transforming repair
tools into feedback generators~\cite{vlhcc17}.
Here, we discuss tools relevant to our work.

\textsc{AutoGrader}~\cite{SGS13} and \Qlose~\cite{qlose}
 can compute 
program repairs 
for incorrect student solutions to introductory programming assignments. 
These systems require the teacher to provide a reference implementation or a comprehensive set of test cases while
\Name also allows students to discover potential repairs using direct manipulations.  
\textsc{AutoGrader}~\cite{SGS13} computes syntactically minimal repairs while 
\Qlose computes syntactically and semantically 
minimal repairs based on a set of test cases.
\Name extends
\Qlose's technique to compute minimal program repairs based on prefix traces and manipulated states.
In particular, \Name encodes the problem of finding good stop points for aligning partial program traces that is novel in our setting of direct manipulation of variables.

\mypar{Program repair}
This topic has been studied extensively
and researchers have proposed techniques based on
constraint-solving \cite{semfix}, 
abstractions \cite{LB12}, and
genetic algorithms~\cite{Arcuri08,LDFW12}.
\Name falls in the first category, but it would be interesting to investigate the effectiveness of the other techniques in our domain.
There are approaches that find repaired programs that are 
syntactically close~\cite{SOE14,directfix} or semantically close~\cite{EssJob13}  to the original program. 
It was demonstrated in \cite{qlose} that repairs generated using a combination of syntactic and semantic program distances 
are, in general, more {\em desirable}. Hence, \Name chooses this last approach.

Existing tools use
test cases~\cite{semfix,qlose}, logic specifications ~\cite{KoukoutosKK16}, or reference programs~\cite{SGS13}.
Direct manipulation ``augments'' a test case by allowing the user to specify intermediate
information about the run of the program on a certain input. Moreover, direct manipulations can be used to debug partially written implementations.
Finally, it is important to note that direct manipulation is not directly expressible
using assertions or test cases: while an assertion at a certain location is valid if \emph{every} time the location
is traversed the predicate in the assertion is true, a direct manipulation at a certain location only requires that 
\emph{at some point} in the trace the variables evaluate to the manipulated values at that location.

Several repair tools use  fault localization to find likely
locations to modify~\cite{BNR03,JM11,CTBB11,KB11}.
Extending fault localization techniques to our setting is another interesting research direction.
The work on angelic debugging~\cite{CTBB11} is particularly relevant to \Name, where the technique infers 
possible faulty expressions in a program by replacing them with an alternate concrete value (oracle) that makes all the tests pass. However, the burden on repairing the program with the correct expression still lies with the programmer.

The \textsc{CeGIS} refinement of external functions presented in Sec.~\ref{subsec:auxiliary} is related to the notion of \textsc{Sketch} models~\cite{modularsketch14},
which allow one to specify certain properties (such as associativity, idempotence, etc.) 
to provide richer interpretations to uninterpreted functions. 
In contrast, $\Name$ iteratively builds a model of the auxiliary function directly from the repair process.

\section{Conclusion}
\label{sec:conclusion}

We presented a new specification mechanism called \emph{direct manipulation},
which allows programmers to  
modify variable values at an intermediate point in a failing trace to indicate what variable values were expected. 
We built the tool \Name that can
repair simple Java programs to satisfy direct manipulations specified by the programmer.
Finally, we showed that direct manipulations lead to more accurate repairs than those
computed using only test cases.
While our approach
has shown promising results on simple programs appearing in introductory programming assignments,
this paper is just the first step towards a new, exciting direction and many research questions remain open.
Our next steps include investigating more complex manipulation techniques that can be used 
by real programmers on large scale programs and
designing user studies to assess whether students/programmers are receptive to this novel
specification paradigm.

%


\bibliographystyle{splncs03}
\bibliography{paper}

\appendix
\section{Repair Model}
\label{sec:repair-model}

Figure~\ref{fig:repairmodel} illustrates \Name's default repair model and Figure~\ref{fig:rewriteexample}
illustrates an example of how the repair model generates a Sketch from a program.
\begin{figure}
\centering
\begin{tabular}{rcl}
\textsc{r}(\t{x = e})  & $\rightarrow$ & x = \textsc{r}(e)\t{ + ($\sum_{\texttt{v}\in V}??_{b}$v+??)}\\
\textsc{r}(\t{e$_1$ bop e$_2$})  & $\rightarrow$ &  \textsc{r}(e$_1$) \t{bop} \textsc{r}(e$_2$)\t{ + ($\sum_{\texttt{v}\in V}??_{b}$v+??)}\\
\textsc{r}(\t{return e})  & $\rightarrow$ &  \textsc{r}(e) \t{ + ($\sum_{\texttt{v}\in V}??_{b}$v+??)}\\
\textsc{r}(\t{e$_1$+e$_2$})  & $\rightarrow$ & \textsc{r}(\t{e$_1$}) + \textsc{r}(\t{e$_2$}) \\
\textsc{r}(\t{c*e}) & $\rightarrow$ & \t{c * }(\textsc{r}(\t{e}) + ($\sum_{\texttt{v}\in V}??_{b}$v+??))\\
\textsc{r}(\t{c}) 	         & $\rightarrow$ & \t{c}\\
\textsc{r}(\t{x}) & $\rightarrow$ & \t{$??_{b}$x}\\
\textsc{r}(\t{x[e]}) & $\rightarrow$ & \t{$??_{b}$x[$??_{b}$*e+??]}\\
\textsc{r}(\t{f(e)}) & $\rightarrow$ & \t{$??_{b}$f(e)}\\
\end{tabular}
\caption{\Name's repair model.
	\label{fig:repairmodel}
	}
\end{figure}

\begin{figure}
\centering
{
		\begin{minipage}{0.3\linewidth}
			\begin{lstlisting}
int triple(int x){
  int y = 3 * x;
  if(x == 10)  
     y = 30;
  return y;
}
			\end{lstlisting}
		\end{minipage}
\hspace{-1mm}			
			$\xrightarrow{\textsc{R}}$
\hspace{-4mm}			
		\begin{minipage}{0.6\linewidth}
			\begin{lstlisting}[mathescape=true]
int SkTriple(int x){
  int y = 3*(??$_b$x+(??$_b$x+??))+(??$_b$*x+??);
  if(??$_b$x == 10 + (??$_b$x+??$_b$y+??)) 
    y = 30 + (??$_b$x+??$_b$y+??);
  return ??$_b$y + (??$_b$x+??$_b$y+??);
}
			\end{lstlisting}
		\end{minipage}
		}
	\caption{A sketched program obtained from 
	applying the repair model to a program.
	Holes of the form  \t{??} can be instantiated with arbitrary integers.
	Holes of the form \t{??$_b$} can only be instantiated with values in $\{-1,0,1\}$.
	}
	\label{fig:rewriteexample}
\end{figure}
\setlength{\tabcolsep}{5pt}

\section{Detailed Benchmark Description}
\label{sec:benchmark}
All benchmarks and the corresponding Sketch files are available at this url:
\url{https://tinyurl.com/yd6bp3dx}.
The set contains three variants of \t{largestGap} (\S~\ref{sec:motivating}), which is used in the Microsoft CodeHunt platform~\cite{codehunt}. 
For two of the \t{largestGap} benchmarks, \t{largestGap-1} and \t{largestGap-3},
we further split the programs into two sub-benchmarks denoted with the suffixes \t{.1} and \t{.2}.
These are benchmarks in which the students has two logically separate errors that would require
many test cases to be fixed. We use \t{.1} to denote the original version of the benchmark and assume
 the task is to fix the first
error in the program and \t{.2} to denote the same benchmark in which the first bug has been fixed and 
the task is to fix the second error.
The \t{tcas-semfix} program is a toy traffic collision avoidance system from~\cite{semfix}. 
The \t{max}, \t{iterPower}, \t{ePoly}, and \t{multIA} problems are taken from the Introduction to Python Programming course taught on edX \cite{edx}.
Two of the new programs we consider are variations of \Qlose benchmarks.
The other three \t{bubbleSort}, \t{subLargestGap}, and \t{maxMin}   are larger programs that contain multiple loops, which are more complex than the benchmarks  considered
  in~\cite{qlose}.

\end{document}